\documentclass[a4paper,twocolumn]{article}

\usepackage[utf8]{inputenc}
\usepackage[T1]{fontenc}

\usepackage{float}
\usepackage{amsmath,amssymb}
\usepackage{mathrsfs}
\usepackage{theorem}
\usepackage{graphicx}
\usepackage{color}
\usepackage{wasysym}
\usepackage{hyperref} 
\usepackage{verbatim}

\definecolor{blue}{rgb}{0,0,1}
\definecolor{green}{rgb}{0,0.65,0.5}
\definecolor{verde}{rgb}{0.,.5,0.4}
\definecolor{marron}{rgb}{0.7,0.2,0.1}
\definecolor{red}{rgb}{1,0,0}
\definecolor{vio}{rgb}{1,0,1}
\definecolor{ama}{rgb}{1,1,0}

\newcommand{\bc}{\begin{center}}
\newcommand{\ec}{\end{center}}
\newcommand{\be}{\nopagebreak[3]\begin{equation}}
\newcommand{\ee}{\end{equation}}
\newcommand{\ba}{\nopagebreak[3]\begin{eqnarray}}
\newcommand{\ea}{\end{eqnarray}}

{\theorembodyfont{\sffamily} }
{\theorembodyfont{\sffamily} }
{\theorembodyfont{\sffamily} }

\begin{document}

\title{\bf 
Constructing balanced equations of motion for
particles in general relativity: \\
the harmonic gauge case
}

\author{
Emanuel Gallo
and
Osvaldo M. Moreschi
\\
{\rm \small Facultad de Matemática Astronomía, Física y Computación (FaMAF),} \\
{\rm \small Universidad Nacional de C\'{o}rdoba,}\\
{\rm \small Instituto de F\'\i{}sica Enrique Gaviola (IFEG), CONICET,}\\
{\rm \small Ciudad Universitaria,(5000) C\'{o}rdoba, Argentina.}
}

\maketitle

\begin{abstract}

Based in the framework of article  \cite{Gallo:2016hpy}, where we have
presented the general problems one encounters in the construction of
balanced equations of motions for particles in relativistic theories
of gravity, 
we present in this work the explicit
balanced equations of motion for a compact object
in general relativity in the harmonic gauge.
\end{abstract}

\vspace{5mm}

\section{Introduction}

The excitement of witnessing this new era of gravitational wave 
observations\cite{Abbott:2016blz,TheLIGOScientific:2016wfe,Abbott:2016nmj,Abbott:2017vtc,
	Abbott:2017oio,TheLIGOScientific:2017qsa,Abbott:2017gyy}
coming from binary black holes and binary neutron stars,
poses also the challenge to have at hand the most convenient models
of the physical system, at every stage of their dynamics;
in order to cope with the description of the data acquisition, which is ever increasing.
In particular we would like to contribute with models that are
useful for the description of the dynamics of coalescence binary systems
and their specific relation with the gravitational emitted radiation.
Our work concentrates in the construction of equations of motion
of compact objects, subjected to the back reaction due
to their emission of gravitational radiation.

In reference \cite{Gallo:2016hpy} we have presented the general
necessary framework to construct balanced equations of motion for particle
in general relativistic theories. 
In this work we present this framework applied to the case of general relativity
in the harmonic gauge.

The target system we have in mind is a binary gravitationally bound isolated
interactive system.
When considering a compact object as isolated, in the framework of isolated spacetimes,
one realizes that in general one can ascribe a flat background to the spacetime.
More concretely, in the asymptotic region one can always write the metric as
\begin{equation}\label{eq:asymp1}
 g = \eta_{\text{asy}} + h_{\text{asy}} 
;
\end{equation}
where $\eta_{\text{asy}}$ is a flat metric associated to inertial frames in the asymptotic
region
 and $h_{\text{asy}}$ the tensor where all the physical information is encoded.
But, there are as many flat metrics $\eta_{\text{asy}}$ as there are BMS\cite{Sachs62,Moreschi86} 
proper supertranslation generators.
And also one knows that the difficulties in finding appropriate rest frames
comes from the existence of gravitational radiation\cite{Moreschi88,Moreschi98,Dain00'}.
At this point it is important to remark that we have shown in the past
that the way to circumvent this problem is to recur to the notion of center of mass
at future null infinity.
The notion of center of mass is in turn related to the definition of angular
momentum. We have solved this problem in \cite{Moreschi04},
where a supertranslation free definition of intrinsic angular momentum
was introduced along with its related center of mass frame.
In other words, for each point at future null infinity, we have a way to single out
a unique decomposition of the metric in the form (\ref{eq:asymp1}), with an
appropriately selected flat background.

It is for this reason that gravitational radiation should be the first quantity
to be taken into account for the discussion of back reaction on the motion
of compact objects. 
Therefore in calculating the appropriate equations of motion for particles
we take this as our starting point; so that the root of the difficulties
is taken care at the beginning of our approach.

We adopt here the viewpoint explained in  \cite{Gallo:2016hpy}
in which we assume there exists an exact metric
$\mathsf{g}$  that corresponds to an isolated 
binary 	system of compact objects;
which it can be decomposed in a form:
\begin{equation}
\mathsf{g}  = \eta + \mathsf{h}_A + \mathsf{h}_B + \mathsf{h}_{AB} ;
\end{equation} 
where $\eta$ is a flat metric,
$ \mathsf{h}_A$ is proportional to a parameter $M_A$, that one can think is some kind
of measure of the mass of system $A$,
similarly $ \mathsf{h}_B$ is proportional to a parameter $M_B$,
and $ \mathsf{h}_{AB}$ is proportional to both parameters.
To study the gravitational radiation emitted by the motion of
particle $A$, 
we model the asymptotic structure of a sub-metric 
\begin{equation}\label{eq:gA}
g_A = \eta + h_A ;
\end{equation}
and to describe the rest of the system, we use
a sub-metric
\begin{equation}\label{eq:gB}
g_B = \eta + h_B .
\end{equation}
For more details see article \cite{Gallo:2016hpy},
where it is also explained that
the appropriate choice of the flat metric $\eta$ 
should be related to a local notion of center of mass frame.

Although we will be studying the dynamics of system $A$, to simplify the 
notation we will avoid using a subindex $A$, whenever possible.

We present in the next section the problematic of the interior region.
Then, in section \ref{sec:exterior} we present the ingredients
of the exterior asymptotic problem.
A discussion of the relation between these two topics is presented
in section \ref{sec:link}.
The balanced equations of motion calculated in the harmonic gauge
are introduced in section \ref{sec:balanced}.
In the last section we make some final remarks on the subject
of this work.
Some auxiliary relations have been included in an Appendix.


\section{The interior problem}\label{sec:interior}

The subject of the interior problem is treated through the approximation based
on the relaxed covariant form of the field equations, as studied in
\cite{Gallo:2012b}.

When solving the relaxed field equations we use as background a flat
background with metric $\eta$.
The relaxed solutions is further restricted to satisfy a dynamical
equation on an auxiliary metric $g_{(B)}$, that takes
into account the existence of the rest of the system, that we call $B$.

We start by presenting a summary of its elements.

\subsection{The decomposition of the metric}
The study of the particle paradigm can be decomposed in an interior problem
and an asymptotic problem. In the interior problem one studies the determination
of the spacetime geometry in terms of the dynamics of the binary system.
The basic assumption takes the total metric as consisting of a flat background
metric plus a correcting terms that has the information of the system.
In what follows we present the basic equations that describe the interior problem
in terms of a general background metric $\tilde \eta$; not necessarily flat.

Let us express the metric $g_{ab}$ of the spacetime $M$
in terms of a reference metric $\tilde \eta_{ab}$, such that
\begin{equation}\label{eq:getamash}
g_{ab} = \tilde \eta_{ab} + h_{ab} ;
\end{equation}
where we use Latin indices to denote abstract indices.

Let $\partial_a$ denote the torsion free metric connection
of $\tilde \eta_{ab}$ and $\nabla_a$ the torsion free metric connection
of $g_{ab}$; then one can express the covariant derivative
of an arbitrary vector $v$ by
\begin{equation}
\nabla_a v^b = \partial_a v^b + \Gamma_{a\;c}^{\;b} v^c ;
\end{equation}
and one can prove that
\begin{equation}
\Gamma_{a\;b}^{\;c}=
\frac{1}{2} g^{cd}
\left(
\partial_a h_{bd} + \partial_b h_{ad} - \partial_d h_{ab}
\right)
= \Gamma_{b\;a}^{\;c} .
\end{equation}

The relation between $\Gamma$ and the curvature tensor
can be calculated from
\begin{equation}
\begin{split}
[\nabla_a,\nabla_b] v^d &=
\Theta_{abc}^{\;\;\;\;\;d} v^c +\\
&\left(
\partial_a \Gamma_{b\;c}^{\;d}
- \partial_b \Gamma_{a\;c}^{\;d}
+
\Gamma_{a\;e}^{\;d} \, \Gamma_{b\;c}^{\;e}
-
\Gamma_{b\;e}^{\;d} \, \Gamma_{a\;c}^{\;e}
\right) v^c \\
&=
R_{abc}^{\;\;\;\;\;d} v^c ;
\end{split}
\end{equation}
where $\Theta$ is the curvature of the $\partial_a$
connection.
Then
the Ricci tensor can be calculated from
\begin{equation}
\begin{split}
R_{ac} & \equiv R_{abc}^{\;\;\;\;\;b} \\
& =
\Theta_{ac} +
\partial_a \Gamma_{b\;c}^{\;b}
- \partial_b \Gamma_{a\;c}^{\;b}
+
\Gamma_{a\;e}^{\;b} \, \Gamma_{b\;c}^{\;e}
-
\Gamma_{b\;e}^{\;b} \, \Gamma_{a\;c}^{\;e} ;
\end{split}
\end{equation}
where $\Theta_{ac}$ is the Ricci tensor of the connexion
$\partial_a$.

\subsection{The field equations in relaxed covariant form}
Let us consider four independent auxiliary functions $x^\mu$,
with $\mu=0,1,2,3$. Then let us observe that
\begin{equation}
g^{ab} \nabla_a \nabla_b x^\mu =
g^{ab} \nabla_a \partial_b x^\mu =
g^{ab} \partial_a \partial_b x^\mu -
g^{ab} \Gamma_{a\;b}^{\;c} \partial_c x^\mu .
\end{equation}
Then, if $I_\mu^{\;e}$ denotes the inverse of $\partial_c x^\mu$,
which exists by assumption of the independence of the set $x^\mu$,
one has
\begin{equation}
g^{ab} \Gamma_{a\;b}^{\;c} =
\left(
g^{ab} \nabla_a \nabla_b x^\mu -
g^{ab} \partial_a \partial_b x^\mu
\right)
I_\mu^{\;c}
=
H^\mu I_\mu^{\;c} ;
\end{equation}
where we are using
\begin{equation}
H^\mu =
-
g^{ab} \nabla_a \nabla_b x^\mu 
+
g^{ab} \partial_a \partial_b x^\mu
 . \label{eq:def-H}
\end{equation}
Alternatively, let us define the gauge vector
\begin{equation}
 \mathscr{H}^c = H^\mu I_\mu^{\;c} ;
\end{equation}
which implies
\begin{equation}
 H^\mu = \partial_c x^\mu \; \mathscr{H}^c =  \mathscr{H}(x^\mu) ;
\end{equation}
then one has
\begin{equation}
g^{ab} \Gamma_{a\;b}^{\;c} = \mathscr{H}^c .
\end{equation}

%
Then the Ricci tensor can be expressed by
\begin{equation}\label{eq:ricciH}
\begin{split}
R_{ac} &= 
\Theta_{ac}
+
\frac{1}{2}
g^{bd}
\left(
 \Theta_{bad}^{\;\;\;\;\;e} h_{ec}
+\Theta_{bcd}^{\;\;\;\;\;e} h_{ea}
+2 \Theta_{bca}^{\;\;\;\;\;e} h_{ed}
\right)
\\
&\quad +
\frac{1}{2} g^{bd} \partial_b \partial_d h_{ac}
-
\partial_{(a}
\left(
g_{c)e} \mathscr{H}^e
\right)
+  g_{ed} \Gamma_{a\;c}^{\;e} \mathscr{H}^d
\\
&\quad -
g^{bf} g_{ed}
\Gamma_{a\;f}^{\;d}
\Gamma_{b\;c}^{\;e}
-
\frac{1}{2}
\left(
\Gamma_{a\;}^{\;bd}
\Gamma_{bcd}
+
\Gamma_{c\;}^{\;bd}
\Gamma_{bad}
\right)
.
\end{split}
\end{equation}

The field equations are
\begin{equation}
\label{eq:eins1}
R_{ac} = -8\pi \kappa
\left(
T_{ac} - \frac{1}{2}g_{ac} g^{bd} T_{bd}
\right) .
\end{equation}

In writing equation (\ref{eq:ricciH}) in a coordinate frame, without any reference to $\tilde \eta$,
one would obtain the analogous expression without the $\Theta$ terms, and where all the
appearance of $\partial$ derivatives are replace by coordinate derivatives.

Alternatively one can use the form of the field equations 
in terms of a slight different logic.

If we use the expression of the Ricci tensor as given by (\ref{eq:ricciH}) in 
(\ref{eq:eins1}), namely
\begin{equation}\label{eq:relaxed}
\begin{split}
&\frac{1}{2} g^{bd} \partial_b \partial_d h_{ac}
-
\partial_{(a}
\left(
g_{c)e} \mathscr{H}^e
\right)
+ g_{ed} \Gamma_{a\;c}^{\;d} \mathscr{H}^e
\\
&
+ \Theta_{ac}
+
\frac{1}{2}
g^{bd}
\left(
 \Theta_{bad}^{\;\;\;\;\;e} h_{ec}
+\Theta_{bcd}^{\;\;\;\;\;e} h_{ea}
+2 \Theta_{bca}^{\;\;\;\;\;e} h_{ed}
\right)
\\
& -
g^{bf} g_{ed}
\Gamma_{a\;f}^{\;d}
\Gamma_{b\;c}^{\;e}
-
\frac{1}{2}
\left(
\Gamma_{a\;}^{\;bd}
\Gamma_{bcd}
+
\Gamma_{c\;}^{\;bd}
\Gamma_{bad}
\right) \\
=&
-8\pi \kappa
\left(
T_{ac} - \frac{1}{2}g_{ac} g^{bd} T_{bd}
\right)
;
\end{split}
\end{equation}
we will refer to this as the \emph{relaxed field equations}\cite{Walker80},
where one has not assumed that $\mathscr{H}^c$ is $g^{ab} \Gamma_{a\;b}^{\;c}$.

Using the standard harmonic gauge technique, one would say:
solve the relaxed field equation in the coordinate frame, with $H^\mu = 0$, and then require the equation
\begin{equation}
\label{eq:armon1}
g^{bd} \nabla_b \nabla_d x^{\mu} =0 .
\end{equation}
In the standard approach one makes use of coordinate basis; therefore the previous statement
would be the complete story. However in our case, $H^\mu$ has a second term where two covariant derivatives 
of $x^\mu$ with respect to the metric $\eta$ appears.
At this point it is important to notice that if we have the solutions $x^\mu$ 
from (\ref{eq:armon1}) then, on constructing $\eta$
with these as a Cartesian coordinate system, one would obtain $H^\mu = 0$.

In some occasions it is preferable to work with a different set of equations.
In this respect, several authors have indicated that actually to request equation (\ref{eq:armon1}) 
is equivalent\cite{Einstein:1938yz,Anderson73,Walker80}
to demand
\begin{equation}
\label{eq:diverg1}
g^{ab} \nabla_a T_{bc} =0 .
\end{equation}
When dealing with Einstein equations in the relaxed form, and treating the vacuum case,
equation (\ref{eq:diverg1}) is understood as the condition that the divergence
of the Einstein tensor must be zero 
(which of course is identically zero in the non relaxed form).

In reference \cite{Gallo:2012b} we have related this approach to the results 
of Friedrich\cite{Friedrich85}.
We have also indicated in this reference how to set up an iterative approximation
scheme to solve the field equations.

The first order correction for the metric is calculated from
\begin{equation}
\label{eq:h1lin}
\frac{1}{2} \eta^{bd} \partial_b \partial_d h^{(1)}_{ac}
=
- 8\pi
\left(
T^{(0)}_{ab} - \frac{1}{2} \eta_{ab} \, \eta^{cd}T^{(0)}_{cd}
\right) ;
\end{equation}
where we are using geometric units.

\subsection{The gravitational field from one particle in first order} 

Let us consider a massive point particle with mass
$M_A$ describing, in a flat space-time $(M^0,\eta_{ab})$, 
a timelike world line curve $C$
which in a Cartesian coordinate system $x^a$ reads
\begin{equation}\label{eq:trajectoria}
 x^\mu =z^\mu(\tau_0),
\end{equation}
with $\tau_0$ meaning the proper time of the particle along $C$.

{The unit tangent vector to $C$, with respect to the flat background metric is}
\begin{equation}\label{eq:4-velocity}
\mathbf{v}^\mu =\frac{dz^\mu}{d\tau_0},
\end{equation} 
that is, 
\begin{equation}\label{eq:vuniteta}
\eta(\mathbf{v} , \mathbf{v})=1 . 
\end{equation}
Now, for each point $Q=z(\tau_0)$ of $C$, we draw a future null cone
$\mathfrak{C}_Q$ with vertex in $Q$. If we call $x^\mu_P$
the Minkowskian coordinates of
an arbitrary
point on the cone $\mathfrak{C}_Q$, then we can define
the retarded radial distance on the null cone by
\begin{equation}\label{eq:retardedr}
r = \mathbf{v}_\mu \left(x^\mu_P-z^\mu(\tau_0) \right).
\end{equation}

The energy momentum tensor $T^{(0)}_{ab}$ of a point particle
is proportional to $m v_a v_b$; where $m$ is the mass
and $v^a$ its four velocity.
{
We are distinguishing between the unit tangent vector $\mathbf{v}^a$,
with respect to the metric $\eta$,
and the four velocity vector $v^a$, because we would like to consider the
possibility to normalize the vector $v$ with respect to a different metric.
}
In order to represent a point particle
$T^{(0)}_{ab}$ must also be proportional to a
three dimensional delta function that has support on
the world line of the particle.

We will assume that the particle does not have multipolar structure.
Then, given an arbitrary Minkowskian frame $(x^0,x^1,x^2,x^3)$,
we will express the
energy momentum by
\begin{equation}\label{eq:tab1}
\begin{split}
&T_{ab}(x^0=z^0(\tau),x^1,x^2,x^3) =
\, M\; v_{a}(\tau)\; v_{b}(\tau) \\
& \frac{\delta(x^1 - z^1(\tau))
\delta(x^2 - z^2(\tau))
\delta(x^3 - z^3(\tau))}
{\sqrt{-\det g^{(3)}} }
;
\end{split}
\end{equation}
where $\tau$ is the chosen parametrization.

Then, 
for a source of the form (\ref{eq:tab1}), independently of the parametrization,
 the solution to the linear field equation is
\begin{equation}\label{eq:h1}
h^{(1)}_{ab} = - 4 M_A\frac{v_a v_b - \frac{1}{2} \eta_{ab}}{r}
 ;
\end{equation}
so that in general
\begin{equation}\label{eq:g1}
g^{(1)}_{(A)\,ab} = \left( 1 + \frac{2 M_A}{r}\right) \eta_{ab}
- \frac{4 M_A}{r} v_a v_b
.
\end{equation}
{
In these equations we have considered the definition
\begin{equation}\label{eqv_a}
 v_a \equiv \eta_{ab} v^b ;
\end{equation}
however it should be emphasized that the vector $v^b$ is not normalized with
the flat metric $\eta$ as we will see below.

}

It is curious that the inverse of this first order metric can be calculated exactly;
namely
\begin{equation}
g^{(1)ab} =  \frac{1}{ 1 + \frac{2 M_A}{r}} \eta^{ab} 
+ \frac{\frac{4 M_A}{r}}{ 1 - \left( \frac{2 M_A}{r}\right)^2 } v^a v^b
;
\end{equation}
if the vector field $v^a$ is normalized with $\eta$, and
\begin{equation}
\begin{split}
g&^{(1)ab} =  \frac{1}{ 1 + \frac{2 M_A}{r}} \eta^{ab} \\
&+ 
\frac{\frac{4 M_A}{r}}
{\big(1 + \frac{2 M_A}{r}\big)
 \big[1 -  \frac{2 M_A}{r}\left(2 \eta(v,v) -1 \right) \big] 
} 
v^a v^b
;
\end{split}
\end{equation}
if $\eta(v,v) \neq 1$.
In any case, at first order these two versions of the inverse coincide.

\subsection{Lowering and raising indices for the velocity vectors 
{and normalization conditions}}

We require
\begin{equation}\label{eq:auno}
\begin{split}
1 =& \, g_{(B)\,ab} v_{(A)}^{\quad a} v_{(A)}^{\quad b} \\
=&
\left( \eta_{ab} + h_{(B)\,ab} \right)
v_{(A)}^{\quad a} v_{(A)}^{\quad b}
,
\end{split}
\end{equation}
and therefore
\begin{equation}\label{eq:etavava}
{v_{(A)b} \; v_{(A)}^{\quad b} = }
\, \eta_{ab} v_{(A)}^{\quad a} v_{(A)}^{\quad b} =
1-h_{(B)\,ab} 
v_{(A)}^{\quad a} v_{(A)}^{\quad b}
.
\end{equation}

Let us note that the two velocity vectors are proportional
\begin{equation}\label{eq:vupsilonv}
 v^b = \Upsilon \mathbf{v}^b .
\end{equation}
So that one has
\begin{equation}
 1 = g_B(v,v) = \Upsilon^2 g_B(\mathbf{v},\mathbf{v})
= \Upsilon^2 \big( 1 +  h_B(\mathbf{v},\mathbf{v}) \big)
 ;
\end{equation}
which gives $\Upsilon$ in terms of $\mathbf{v}$ and $g_B$.

For later reference, we will use $\tau$ to denote the proper the
with respect to metric $g_B$ and $\tau_0$ for the symbol of the proper
time with respect to the metric $\eta$.

\subsection{Local dynamics at zero order in back reaction}

The expression for the momentum and the local dynamics at zero order
has been discussed in \cite{Gallo:2016hpy}.
We here just mentioned the fundamental expressions.

Let $e_{\underline{c} }^{\;\; b}$ be a frame basis
(an orthonormal frame with respect to metric $g_{(B)}$);
where, as before, 
we are distinguishing between the abstract indices $(a,b,c,..)$ and the numeric frame indices
$(\underline{a},\underline{b},\underline{c},...)$ and let $\theta^{\underline{c}}_{\;\;b}$
be its coframe.
Then the local notion of momentum for particle $A$ is given by
\begin{equation}
 P_{(A)\underline{c}} = {M_A} g_{(B)}(v_{(A)}, e_{\underline{c} })
.
\end{equation}

The local dynamics,
at zero order in back reaction due to gravitational radiation, 
is determined by
\begin{equation}
\begin{split}
M_A \, g_{(B)ab}
\left(
v_{(A)}^{\quad e} \nabla_{(B)e}(v_{(A)}^{\quad a}) e_{\underline{c} }^{\; b}
\right)
= 0
.
\end{split}
\end{equation}
When considering below the effects of back reaction, the right hand side
will turn different from zero.

\section{The exterior asymptotic \\
 problem}\label{sec:exterior}

\subsection{Total momentum}

Given any section $S$ at future null infinity, the total momentum
of a generic spacetime, in terms of an
inertial (Bondi) frame\cite{Moreschi86}, is normally given by
\begin{equation}\label{eq:totalp}
\mathcal{P}^{\mu} =  - \frac{1}{4 \pi} \int_S \hat l_0^{\mu} 
(\Psi_2^0 + \sigma_0 \dot {\bar \sigma}_0) dS^2 ,
\end{equation}
where
the auxiliary null vector 
$\hat l_0(x^2,x^3)$, is defined
in terms of the angular coordinates $(x^2,x^3)$,
by
\begin{equation}  \label{eq:bflhat}
\begin{split}
\hat l_0^\nu(x^2,x^3)
\equiv &
 \biggl(1,\sin(\theta) \cos(\phi), \sin(\theta) \sin(\phi), \cos(\theta) \biggr)\\
=&\left( 1,\frac{\zeta +\bar \zeta }{
1+\zeta \bar{\zeta }},\frac{\zeta -\bar{\zeta }}{i(1+\zeta
\bar{\zeta )}},\frac{\zeta \bar{\zeta }-1}{1+\zeta \bar{\zeta
}}\right);
\end{split}
\end{equation}
where $\mu, \nu, \cdots =0,1,2,3$,
and we are using either the standard sphere angular coordinates 
$(\theta,\phi)$ or the complex angular coordinates $\zeta =\frac{\hat x^2 + i \hat x^3}{2}$,
where $(\zeta,\bar\zeta)$ are complex stereographic coordinates
of the sphere; which are related to the standard coordinates by
$\zeta = e^{i \phi} \cot(\frac{\theta}{2})$;
and here a dot means $\frac{\partial}{\partial \tilde u}$; i.e. the partial derivative with respect
to the inertial asymptotic time $\tilde u$, $\Psi_2^0$ is a component of the Weyl tensor in the 
GHP\cite{Geroch73} notation and $\sigma_0$ is the leading order behavior of the spin
coefficient $\sigma$ in terms of the asymptotic coordinate $\hat r$.
So the set of intrinsic inertial coordinates are $(\tilde u,\theta,\phi)$ or
$(\tilde u,\zeta,\bar\zeta)$.

\subsection{The total momentum and flux from charge integrals}
The total momentum for the monopole particle can be calculated using the
charge integral $Q_S$ of the Riemann tensor technique as explained in \cite{Gallo:2016hpy}.
In this subsection we will use the notation of reference \cite{Moreschi04} to reproduce
the essential equations of our first work.

Using the equations of reference \cite{Geroch73} one can obtain
expressions for the asymptotic fields with respect to an inertial
reference frame:
the radiation field
\begin{equation}
 \Psi_3 = \frac{\eth_0 \sigma'_0}{r^2} + \mathscr{O}(\frac{1}{r^3}) ,
\end{equation}
the relation between the leading order behavior of the shear and primed shear
\begin{equation}
 \dot \sigma_0 = - \bar \sigma'_0 ,
\end{equation}
the time derivative of the leading order behavior of $\Psi_2$
\begin{equation}
 \dot \Psi_2^0 = \eth_0 \Psi_3^0 + \sigma_0 \Psi_4 ,
\end{equation}
and the leading order behavior of the radiation component
\begin{equation}\label{eq:radiationPsi4}
 \Psi_4^0 = \dot \sigma'_0 .
\end{equation}

In the calculation for the flux of the momentum one takes $w_2=0$,
and obtains\cite{Gallo:2016hpy}:
\begin{equation}
\begin{split}
 \dot Q_S(w) 
=& 4 \int  \left[
2  w_1 ( \bar\sigma'_0 \sigma'_0 )
\right] dS^2 + {\tt c.c.} 
\end{split}
\end{equation}
Therefore we can encode the radiation flux of momentum
in terms of the $(\sigma'_0 \bar \sigma'_0)$ factor;
which is gauge invariant as discussed in \cite{Gallo:2016hpy}.



It follows that
the time variation of the Bondi momentum, {of a generic spacetime,} 
can be expressed by
\begin{equation}\label{eq:bondibalance}
\dot{\mathcal{P}}^{\mu} =  - \frac{1}{4 \pi } \int_S 
\hat l_0^{\mu} \, \sigma'_0 \, {\bar \sigma}'_0 \, dS^2 
\equiv -\mathcal{F}^{\mu}
;
\end{equation}
that is, $\mathcal{F}^{\mu}$ is the total momentum flux.

Normally one expresses the dynamics in terms of the proper time of the particle.
As explained in reference \cite{Gallo:2016hpy} the relation between the
asymptotic inertial time $\tilde{u}$ and the dynamical time $u$
can be expressed as
\begin{equation}\label{eq:vtilde}
 \frac{\partial \tilde u}{\partial u} 
= \tilde V(u,\zeta,\bar\zeta) > 0 ,
\end{equation}
that is, in terms of
the time derivative of the inertial time $\tilde u$ 
with respect to the non-inertial time $u$.
This in turn introduces the new information in the 
scalar $\tilde V(u,\zeta,\bar\zeta)$; which will appear in
the new expressions.

Then the time derivative of the total momentum with respect to the
new dynamical time is given by
\begin{equation}\label{eq:bondibalanceV}
\frac{d\mathcal{P}^{\mu} }{du}
 =  - \frac{1}{4 \pi } \int_S 
\hat l_0^{\mu} \tilde V \, \sigma'_0 \, {\bar \sigma}'_0 \, dS^2 
= -\mathcal{F}_V^{\mu}
;
\end{equation}
where now $\mathcal{F}_V^{\mu}$ is the instantaneous momentum flux
with respect to the time $u$.
Therefore, when considering non-inertial times, one only
needs to calculate the radiation scalar $\sigma'_0$ to evaluate the flux of gravitational
radiation.

\section{The link between the interior and asymptotic problem}\label{sec:link}

\subsection{Preliminary considerations}
It is appropriate at this point to recall the experience gained in the case
of charged particles in Minkowski spacetime.
Forcing the equation of motion to balance the electromagnetic radiation
we have derived\cite{Gallo:2011tf} a new general equation of motion that in a particular case
gives well known Lorentz-Dirac equations of motion.

We will now apply the notion of balance equations of motion to the case of
a binary system in general relativity.

So the technique is as follows. We know that the asymptotic equations imply
the balance of Bondi total momentum with the total flux of momentum at future
null infinity. We use this as a tool to find the correct correction to the
zero order equations of motion that would satisfy this balance equation.

Also, let us note that the equations must be local, in the sense that the corrections
must be expressed in terms of local quantities describing the geometry
and the motion.
These restrictions leads us to a natural first order correction
to the equations of motion that we describe below.

\subsection{The Asymptotic structure}

We have indicated above
that in these kind of approximations one should not focus on the 
concept of spin coefficient $\sigma$, but instead should consider as more
representative of the radiation content the spin coefficient $\sigma'$;
since, independent of the gauges one always have
\begin{equation}\label{eq:sigma'0a}
 \sigma'_0 = \frac{1}{2} \frac{\partial h_{0\bar m \bar m}}{\partial \tilde u} ;
\end{equation}
and
\begin{equation}
 \Psi_4^0 = \frac{\partial \sigma'_0}{\partial \tilde u} ;
\end{equation}
where $\tilde u$ is an asymptotic inertial null retarded coordinate.

Let us review here the set of dynamical times one could use.
Several of the asymptotic equations are presented in terms of an inertial
frame, which uses the inertial asymptotic time $\tilde u$.
However, since the interior dynamics is either presented in terms of
the proper times $\tau$ or $\tau_0$, other null coordinates
naturally appear that we mention next.

The asymptotic null function $u$ is defined so that at zero order
in the strength of gravitational radiation  $\lambda$\cite{Gallo:2016hpy}
it agrees with the proper time, i.e. $u=\tau$.
However, when gravitational radiation effects are considered then
$\frac{du}{d\tau}$ becomes a new degree of freedom of the problem.

Among the two natural dynamical times we have the relation
$\Upsilon \frac{d}{d\tau_0} = \frac{d}{d\tau}$ (see equation (\ref{eq:vupsilonv}) above.).

The tensor $h$ for a one particle system is:
\begin{equation}\tag{\ref{eq:h1}}
h^{(1)}_{ab} = - 4 M_A\frac{v_a v_b - \frac{1}{2} \eta_{ab}}{r} .
\end{equation}
So that the components with respect to the inertial null tetrad vectors are:
\begin{align}
h_{ll} &= h_{ab} \hat l^a \hat l^b = - \frac{4 M_A}{r} \Upsilon^2 V_\eta^2 , \\
h_{lm} &= - \frac{4 M_A}{r} \Upsilon^2 V_\eta \eth_0 V_\eta , \\
h_{ln} &= - \frac{2 M_A}{r}  \bigg( \Upsilon^2 V_\eta V_{\eta-}  - 1 \bigg) , \\
h_{mm} &= - \frac{4 M_A}{r} \Upsilon^2  (\eth_0 V_\eta)^2   , \\
h_{m \bar m} &= - \frac{4 M_A}{r} \Big(\Upsilon^2 \bar\eth_0 V_\eta \eth_0 V_\eta 
                       + \frac{1}{2} \Big) , \\
h_{nm} &= - \frac{2 M_A}{r} \Upsilon^2 V_{\eta-} \eth_0 V_\eta   , \\
h_{nn} &= - \frac{ M_A}{r} \Upsilon^2 V_{\eta-}^2 ;
\end{align}
and the complex conjugate deduced from them.
The details of the notation are described in the appendix.
However we should also remark that the asymptotic radial coordinate $r$ is
defined with respect to the intrinsic system; and that one has 
the relation $r = V_\eta \hat r + \mathscr{O}(\hat r^0)$;
so that
\begin{equation}
 \frac{1}{2}  h_{0\bar m \bar m} 
=
- 2 M \frac{\Upsilon^2}{V_\eta} (\bar\eth_0 V_\eta)^2
;
\end{equation}
and therefore
\begin{equation}\label{eq:sigmap0}
\begin{split}
 \sigma'_0 =  -4 M_A 
\Big(&
\frac{\Upsilon^2}{V_\eta}  \bar\eth_0 \tilde{\dot V}_\eta \, \bar\eth_0 V_\eta 
+ 
\frac{\Upsilon}{V_\eta} \tilde{\dot\Upsilon}  \big(\bar\eth_0  V_\eta\big)^2  \\
&-
\frac{1}{2}
\tilde{\dot V}_\eta 
\frac{\Upsilon^2}{V_\eta^2} \, (\bar\eth_0 V_\eta)^2
\Big)
;
\end{split}
\end{equation}
where a tilde dot means $\frac{\partial}{\partial \tilde u}$;
i.e. derivative with respect to the asymptotic inertial time $\tilde u$.


Relating this calculation of the radiation field with the general discussion of balanced
equations of motion presented in our previous article, it is clear that this expression
should be understood as being evaluated at the retarded time $u$
with respect to the first order metric $g_A$; although actually it has been
evaluated at the flat background retarded time.
Then, taking into account the general relation (\ref{eq:vtilde}) one can also express
\begin{equation}
 \tilde{\dot V}_\eta
=
\frac{\dot V_\eta}{\tilde V }
=
\frac{du}{d\tau} \frac{\dot V_\eta}{ V }
=
\frac{\frac{d V_\eta}{d\tau} }{ V }
;
\end{equation}
where now a dot means $\frac{\partial}{\partial u}$;
i.e. derivative with respect to the asymptotic intrinsic time $u$,
and
with the new notation
\begin{equation}\label{eq:vee}
 V =  \frac{\partial \tilde u}{\partial \tau} 
.
\end{equation}
Employing the proper time $\tau_0$, one can also express
\begin{equation}
 \tilde{\dot V}_\eta
=
\frac{\frac{d V_\eta}{d\tau_0} }{ V_0 }
;
\end{equation}
where we are using the notation
\begin{equation}\label{eq:vee}
 V_0 =  \frac{\partial \tilde u}{\partial \tau_0} 
.
\end{equation}
Similarly, we can express
\begin{equation}
 \tilde{\dot\Upsilon}
=
\frac{\frac{d \Upsilon}{d\tau_0} }{ V_0 }
.
\end{equation}

Then, one can write the radiation field in terms of $\tau_0$ as
\begin{equation}\label{eq:sigmap0_tau0}
\begin{split}
 \sigma'_0 =  -4 M_A 
\bigg(&
\frac{\Upsilon^2}{V_\eta}
\bar\eth_0 \Big( \frac{\frac{d V_\eta}{d\tau_0} }{ V_0 } \Big)
 \, \bar\eth_0 V_\eta 
+ \frac{\Upsilon}{V_\eta}
\frac{\frac{d \Upsilon}{d\tau_0} }{ V_0 }
\big(\bar\eth_0  V_\eta\big)^2  \\
&-
\frac{1}{2}
\frac{\frac{d V_\eta}{d\tau_0} }{ V_0 }
\frac{\Upsilon^2}{V_\eta^2} \, (\bar\eth_0 V_\eta)^2
\bigg)
.
\end{split}
\end{equation}
In this expression it should be observed that
$V_0 = V_\eta + \mathscr{O}(\lambda)$;
since when no gravitational radiation effects are present,
the structure of the asymptotic spacetime agrees with that of
an stationary spacetime, and so there is one preferred asymptotic
flat background metric, and in this situation one chooses the 
interior flat background metric to agree with the asymptotic one;
which coincides with the center of mass reference frame in both
regimes.
In this way
translations in the interior would agree with translations
in the asymptotic region.
Then since, the radiation field $\sigma'_0$ is by definition order $\lambda$;
we can use at first order in gravitational radiation the expression
\begin{equation}\label{eq:sigmap0_tau0_o1}
\begin{split}
 \sigma'_0 =  -4 M_A 
\bigg(&
\frac{\Upsilon^2}{V_\eta}
\bar\eth_0 \Big( \frac{\frac{d V_\eta}{d\tau_0} }{ V_\eta } \Big)
 \, \bar\eth_0 V_\eta 
+ \frac{\Upsilon}{V_\eta}
\frac{\frac{d \Upsilon}{d\tau_0} }{ V_\eta }
\big(\bar\eth_0  V_\eta\big)^2  \\
&-
\frac{1}{2}
\frac{\frac{d V_\eta}{d\tau_0} }{ V_\eta }
\frac{\Upsilon^2}{V_\eta^2} \, (\bar\eth_0 V_\eta)^2
\bigg)
.
\end{split}
\end{equation}
This is the expression that is useful for further calculations
of the equation of motion, since it is already in terms of the
flat proper time, and Minkowskian quantities.

\section{Balanced particle dynamics}\label{sec:balanced}

Following our previous work, we define the flux vector 
with respect to the proper time $\tau_0$ as:
\begin{equation}\label{eq:force_tau_0}
 \mathbf{F}_0^\mu 
 = - \frac{1}{4 \pi } \int_S 
{\hat l}_0^{\mu} \ V_0 \, \sigma'_0 \, {\bar \sigma}'_0 \, dS^2 
;
\end{equation}
where we use the definition (\ref{eq:vee}).
So that at lowest order  we can just express it as:
\begin{equation}\label{eq:force_tau_0_final}
\mathbf{F}_0^\mu 
= - \frac{1}{4 \pi } \int_S 
{\hat l}_0^{\mu} \ V_\eta \, \sigma'_0 \, {\bar \sigma}'_0 \, dS^2 
.
\end{equation}

Then, as explained in our previous work, the equation of motion in the final form is:
\begin{equation}\label{eq:balanced-int_0}
M_A \bigg(
\mathbf{v}^a \partial_a \, \mathbf{v}^b 
+  \gamma^{\;b}_{a\;\;c} \, \mathbf{v}^a \,  \mathbf{v}^c
+ 
\Big(
\frac{1}{\Upsilon}  \frac{d\Upsilon}{d\tau_0}
+
\frac{w}{\Upsilon}
\Big)
\, \mathbf{v}^b
\bigg)
=
\frac{1}{\chi\Upsilon}  \mathbf{F}_0^\mu
.
\end{equation}
when expressed in terms of the $\tau_0$ dynamical time.

Contracting this equation with $\eta_{bd} \mathbf{v}^d$ gives
\begin{equation}\label{eq:balanced-int_0-con-v}
\begin{split}
M_A \Big(&
\gamma^{\;b}_{a\;\;c} \, \mathbf{v}^a 
\,  \mathbf{v}^c \, \eta_{bd} \mathbf{v}^d
+
\frac{1}{\Upsilon}  \frac{d\Upsilon}{d\tau_0}
+
\frac{w}{\Upsilon}
\Big) 
=
\frac{1}{\chi\Upsilon}
\, \mathbf{F}_0^b \eta_{bd} \mathbf{v}^d
;
\end{split}
\end{equation}
and it remains the equation of motion
\begin{equation}\label{eq:balanced-int_0-flat}
\begin{split}
M_A \,
\mathbf{a}^b
= M_A \, \mathbf{f}_\perp^b
+ 
\frac{1}{\chi\Upsilon}
\, \mathbf{F}_0^d \;
\big(
\eta_d^{\;\;b} - \mathbf{v}_d \mathbf{v}^b
\big)
;
\end{split}
\end{equation}
where we are using the notation:
\begin{equation}
\mathbf{a}^a \equiv \mathbf{v}^b \partial_b \mathbf{v}^a ,
\end{equation}
and
\begin{equation}
\mathbf{f}_\perp^b \equiv - \gamma^{\;d}_{a\;\;c} \, \mathbf{v}^a \,  \mathbf{v}^c 
\big(
\eta_d^{\;\;b} - \mathbf{v}_d \mathbf{v}^b
\big)
;
\end{equation}
which, it should be remarked, only depends on the background 
geometry $g_B$ and $\mathbf{v}$.

Equation (\ref{eq:balanced-int_0-con-v}) is understood as an equation for $w$.
The main dynamical equation is then (\ref{eq:balanced-int_0-flat});
where the possible degree of freedom $\chi$ depends on the detail
nature of gauge being used.
In our case, if we call $u'$ the asymptotic coordinate associated to the
interior time $\tau_0$; then, in the isolated case, 
in the harmonic gauge, at the first order,
a null function $u^*$ will be asymptotically defined by a relation
of the form $u'= u^* + r^*(r)$, as is known from the Schwarzschild case.
But what is important in our case is that asymptotically one
has $du' = du^*$, at fixed $r$. 
It is reasonable to assume this behavior in this setting, 
so that we take $\chi=1$ in our model.

Let us note that these balanced equations of motion indicate
that the back reaction due to gravitational radiation is
described by the force $\mathbf{F}_0^d$, which in turn is determined
by $\sigma'_0$, that is given in terms of the mechanical information
of the particle.
So, although the equations of motion are completely specified,
they are complicated enough
that we plan to calculate them numerically;
when applying the model to specific systems.

\section{Final comments}

In this work we have presented the equations of motion for particles
that take into account the first order back reaction due to gravitational
radiation, using the framework of general relativity, with 
Hilbert-Einstein field equations, in the harmonic gauge.
We have used in this work for the submetric $g_A$ the structure
that comes from the first order iteration of the field equations;
but our work can be extended to higher orders.
This is the natural first calculation of the balanced equations of
motion approach, we have presented in \cite{Gallo:2016hpy};
since most of the literature dealing with equations of motion
use the harmonic gauge, as for example in classical post-Newtonian
works.
In order to build a model with grater predicting power, we need
to develop our equations of motion to higher orders.
The complicated extension to higher order calculations of our
work will be presented in the future.

One of the points to be remarked is the identification between the interior
center of mass flat background metric, with the asymptotic center of mass
flat metric in the asymptotic region.
This plays a central conceptual role in our approach; but this point
is completely missing in other works on the subject of equations
of motion.

We plan to apply these equations of motion to a variety of physical
systems, including the observed gravitational wave data.

It is a priori difficult to advance in what regime our approach
should or should not compare with previous ones as for
example post-Newtonian approaches; since the starting points
are completely different, and our model is designed to
deal at first order with back reaction of gravitational
radiation. 
Also, because our equations are still relativistic,
in terms of individual proper times; instead of the 
universal time approach for the post-Newtonian works.
In \cite{Moreschi:2018gmf} we have presented a technique to calculate retarded
effects on flat background to arbitrary precision; that we intend to apply
to this model. 
This is another difference with the classical post-Newtonian approach,
in which the calculation of the retarded effects are related to
the particular order of the approximation.
Before applying our model to specific cases, it is difficult to
forecast a qualitative comparison with the classical post-Newtonian approach.
But in any  case,
we would like to study the comparison in detail.

In a future work we will present the corresponding balanced equations of motion
in general relativity in the null gauge; where it will be clear that
the approach changes completely, but that all the general considerations
presented in  \cite{Gallo:2016hpy} also apply.

\subsection*{Acknowledgments}

We acknowledge support from CONICET, SeCyT-UNC and Foncyt.

\appendix

\section{Interior reference frames at our disposal}

In what follows we mention some reference frames and coordinates we have
at our disposal to carry out the calculation of the back reaction to the particle
due to the gravitational radiation emitted.

\subsection{Relations among coordinates and null 
vectors in the flat background}

\vspace{2mm}
\noindent
{\bf The inertial system}
\vspace{1mm}

Let us denote with  $y^\mu$  the standard Cartesian coordinates and with  
$(\hat x^0 = \hat u,\hat x^1 = \hat r,\hat x^2,\hat x^3)$,
where $(\hat x^2,\hat x^3)=(\theta,\phi)$ or $\zeta =\frac{\hat x^2 + i \hat x^3}{2}$,
the corresponding null polar coordinates; then, the relation 
between them is given by
\begin{equation}\label{eq:nulpolar}
y^\mu = \hat u \, \delta^\mu_0 + \hat r \, \hat l^\mu(\zeta, \bar \zeta )
;
\end{equation}
with
\begin{equation}
\hat l^\mu(\zeta, \bar \zeta ) = \hat l_0^\mu(\zeta, \bar \zeta ) ;
\end{equation}
defined in (\ref{eq:bflhat}).

\vspace{2mm}
\noindent
{\bf The intrinsic non-inertial system}
\vspace{1mm}

Let $z^\mu(\tau_0)$ be the evolution of the particle with proper time $\tau_0$.
We define a null function $u'$ as the future null cones emanating from
$z^\mu(\tau_0)$, such that $u'= \tau_0$ at the world line of the particle.

If  $(x^0\! =\! u',x^1\! =\! r, x^2, x^3)$,
where $(x^2,x^3)=(\theta',\phi')$ or $\zeta' =\frac{x^2 + i x^3}{2}$,
are the null polar coordinates adapted to an arbitrary 
timelike curve, determined by  $z(u')^\mu$, then one has
\begin{equation}\label{eq:nulpolar'}
y^\mu= z^\mu(u')+r \, \mathbf{l}^\mu(u,\zeta', \bar \zeta' ) 
.
\end{equation}

Note that
\begin{equation}
(y^\mu  - z^\mu(u')) \mathbf{l}_\mu  = r \, \mathbf{l}^\mu \mathbf{l}_\mu = 0 ,
\end{equation}
and that
\begin{equation}
(y^\mu  - z^\mu(u')) \mathbf{v}_\mu  = r \, \mathbf{l}^\mu \mathbf{v}_\mu = r ;
\end{equation}
so that
\begin{equation}\label{eq:boldl}
\mathbf{l}^\mu = \frac{y^\mu  - z^\mu(u')}{(y^\nu  - z^\nu(u')) \mathbf{v}_\nu} .
\end{equation}

Note also that
\begin{equation}
\begin{split}
\frac{\partial r}{\partial y^\nu} 
=&
\delta^\mu_\nu  \mathbf{v}_\mu 
- \mathbf{v}^\mu \frac{\partial u'}{\partial y^\nu} \mathbf{v}_\mu 
+ (y^\mu  - z^\mu(u')) \mathbf{a}_\mu \frac{\partial u'}{\partial y^\nu} \\
=&
\mathbf{v}_\nu 
+ \left( (y^\mu  - z^\mu(u')) \mathbf{a}_\mu -1 \right) \frac{\partial u'}{\partial y^\nu} .
\end{split}
\end{equation}
Also we have
\begin{equation}
\delta_\nu^\mu = \mathbf{v}^\mu \frac{\partial u'}{\partial y^\nu}
+ \frac{\partial r}{\partial y^\nu} \mathbf{l}^\mu
+ r \frac{\partial \mathbf{l}^\mu}{\partial y^\nu} ;
\end{equation}
which implies
\begin{equation}\label{eq:gradu}
\begin{split}
\mathbf{l}_\mu \delta_\nu^\mu 
= \mathbf{l}_\nu
=&
\frac{\partial u'}{\partial y^\nu}
+ r\mathbf{l}_\mu \frac{\partial \mathbf{l}^\mu}{\partial y^\nu} \\
=&
\frac{\partial u'}{\partial y^\nu}
.
\end{split}
\end{equation}

Therefore we have
\begin{equation}\tag{\ref{eq:gradu}}
\frac{\partial u'}{\partial y^\nu} = \mathbf{l}_\nu ,
\end{equation}
and
\begin{equation}\label{eq:gradr}
\frac{\partial r}{\partial y^\nu} 
=
\mathbf{v}_\nu 
+ \left( (y^\mu  - z^\mu(u')) \mathbf{a}_\mu -1 \right) \mathbf{l}_\nu ;
\end{equation}
which are needed for the derivatives of $h_{ab}$ with respect to 
intrinsic coordinates.

Given a fixed point $y^\mu$ one has to take different spacelike directions,
and therefore different angular coordinates, for the two null vectors
to reach the fixed point.
But, if given a particular future null cone determined by the apex $z(u')$,
one also chooses an inertial frame with origin at this apex, then, from
equations (\ref{eq:nulpolar}) and (\ref{eq:nulpolar'}) one deduces that
at this cone the two null vectors $\mathbf{l}^\mu(u,\zeta', \bar\zeta')$
and $\hat l_0^\mu(\zeta', \bar\zeta')$ must be proportional;
so that
\begin{equation}
\mathbf{l}^\mu(u,\zeta', \bar \zeta' ) 
= \alpha(u,\zeta', \bar \zeta' ) \hat l_0^\mu(\zeta', \bar \zeta' ) ;
\end{equation}
with $\alpha>0$.
But then
we have
\begin{equation}
1 = \mathbf{l}^\mu \mathbf{v}_\mu 
= \alpha \hat l_0^\mu \mathbf{v}_\mu
;
\end{equation}
which implies that
\begin{equation}\label{eq:v-eta-1}
\frac{1}{\alpha}  
=  V_\eta
,
\end{equation}
with
\begin{equation}\label{eq:v-eta}
V_\eta \equiv  \hat l_0^\mu \mathbf{v}_\mu
,
\end{equation}
and also
\begin{equation}\label{eq:lconlhat}
\mathbf{l}^\mu(u,\zeta',\bar\zeta') 
= \frac{1}{V_\eta(u,\zeta',\bar\zeta')} \hat l_0^\mu(\zeta',\bar\zeta')
.
\end{equation}

\subsection{The other null tetrad vectors in the flat background}
{\bf The inertial frame}

Let us define the complex vector
\begin{equation}\label{eq:mhat}
\hat m = \frac{\sqrt{2} P_0}{\hat r} \frac{\partial}{\partial \zeta}
,
\end{equation}
and let us act on the Cartesian coordinates; so that
\begin{equation}
\hat m(y^\mu)
= \hat m^\mu
=
\hat m( \hat r \hat l^\mu )
=
\hat r \hat m( \hat l^\mu )
=
\hat r  \frac{\sqrt{2} P_0}{\hat r} \frac{\partial \hat l^\mu}{\partial \zeta}
= \eth_0( \hat l^\mu )
;
\end{equation}
where we recall that $\eth_0$ is the edth operator in the GHP\cite{Geroch73} notation
for the unit sphere.
Therefore, we see that the definition (\ref{eq:mhat}) agrees with
the simple recipe
\begin{equation}
\hat m^\mu(\zeta, \bar\zeta)
= \eth_0\big( \hat l^\mu(\zeta, \bar\zeta) \big)
.
\end{equation}
Let us also note that
\begin{equation}
0= \eth_0( \hat l^\mu \hat l_\mu ) 
= 2\eth_0( \hat l^\mu) \hat l_\mu 
;
\end{equation}
so that it agrees with the orthogonality condition between $\hat l$
and $\hat m$.

It remains to see if one could express the other null tetrad vector $\hat n$
also in terms of the initial expression for $\hat l^\mu$.
Let us consider an expression of the form
\begin{equation}
\hat n^\mu(\zeta, \bar\zeta) = \alpha \hat l^\mu(\zeta, \bar\zeta) 
+ \beta \bar\eth_0 \eth_0 \hat l^\mu(\zeta, \bar\zeta)
;
\end{equation}
for local coefficients $\alpha$ and $\beta$ that we set next.

Let us recall that the operator edth satisfies the property
\begin{equation}
\bar\eth_0 \eth_0 Y_{lm} = -\frac{l(l+1)}{2} Y_{lm}
;
\end{equation}
so that
\begin{equation}
\bar\eth_0 \eth_0 \hat l^i = - \hat l^i
;
\end{equation}
for $i=1,2,3$.
Then since,
\begin{equation}
\sum_i (\hat l^i)^2 = 1 ;
\end{equation}
one has
\begin{equation}
\hat l_\mu  \bar\eth_0 \eth_0   \hat l^\mu = 1
;
\end{equation}
which indicates that one must take $\beta = 1$.
Then, the positive $\alpha$ is set by the condition that $\hat n$ must be null;
so that
\begin{equation}
0 = 2 \alpha \beta \hat l_\mu  \bar\eth_0 \eth_0   \hat l^\mu 
+ \beta^2 (- \sum_i (\hat l^i)^2 )
= 2 \alpha - 1 ;
\end{equation}
so that
\begin{equation}
\hat n^\mu(\zeta, \bar\zeta) = 
\frac{1}{2} \hat l^\mu(\zeta, \bar\zeta) 
+  \bar\eth_0 \eth_0 \hat l^\mu(\zeta, \bar\zeta)
.
\end{equation}

It is convenient at this point to note the difference between the scalar
\begin{equation}\tag{\ref{eq:v-eta}}
V_\eta = \hat l^\mu \mathbf{v}_\mu ,
\end{equation}
and
\begin{equation}\label{eq:ve}
V_M \equiv \hat l^\mu {v(\tau)}_\mu 
;
\end{equation}
with $\tau$ the proper time with respect to the first order background metric
$g_B$, for the case of the binary system $(A,B)$.
Let us emphasize that the velocity vectors are proportional,
as indicated in (\ref{eq:vupsilonv}),
so that
\begin{equation}\label{eq:vconveta}
V_M = \Upsilon V_\eta 
.
\end{equation}

We see then that
\begin{equation}\label{eq:vehatm}
v_\mu \hat m^\mu = \eth_0 V_M ;
\end{equation}
while
\begin{equation}\label{eq:vehatn}
v_\mu \hat n^\mu = \frac{1}{2} V_- ;
\end{equation}
where $V_- = V_M(-v^i)$; that is, it is the corresponding $V_M$ for which the 
spacelike components $v^i$ of the velocity $v^\mu$ have been replaced by
their negative values.
We also define the corresponding $V_{\eta-} = V_\eta(-v^i)$, so that one can also
write
\begin{equation}\label{eq:vehatn2}
v_\mu \hat n^\mu = \frac{\Upsilon}{2} V_{\eta-} .
\end{equation}

Let $A^a$ and $B^a$ be any two vector fields that do not depend on the angular variables, 
and let us define $A = \hat l^\mu A_\mu$
and $B = \hat l^\mu B_\mu$. Then, calculating the scalar product of the two vectors
one obtains
\begin{equation}\label{eq:escalar}
\begin{split}
\eta_{ab} A^a B^b 
=&
\big( \hat l^\mu \hat n^\nu + \hat n^\mu \hat l^\nu
- \hat m^\mu \bar{\hat m}^\nu - \hat m^\mu \bar{\hat m}^\nu \big)
A_\mu B_\nu \\
=&
A B + A \, \bar\eth_0 \eth_0 B  + B \, \bar\eth_0 \eth_0 A
- \eth_0 A \, \bar\eth_0 B\\ 
& - \bar\eth_0 A \, \eth_0 B ;
\end{split}
\end{equation}
which is a very useful equation that is used frequently; as for example
in the components of $h_{ab}$ in terms of the null tetrad base.
In particular we have
\begin{equation}
1 = V_\eta^2  + 2 V_\eta \, \bar\eth_0 \eth_0 V_\eta - 2 \eth_0 V_\eta \, \bar\eth_0 V_\eta 
,
\end{equation}
and
\begin{equation}
\begin{split}
\Upsilon^2
=& \, V_M^2  + 2\, V_M \, \bar\eth_0 \eth_0 V_M - 2\, \eth_0 V_M \, \bar\eth_0 V_M \\
=& \, V_M \, V_- - 2 \, \eth_0 V_M \, \bar\eth_0 V_M
;
\end{split}
\end{equation}
from which
\begin{equation}
\bar\eth_0 \eth_0 V_M 
=
\frac{1}{2} \big( V_- - \, V_M \big)
;
\end{equation}
without forgetting (\ref{eq:vconveta}).

%
%
%

\end{document}